\documentclass{article}
\usepackage{ltexpprt,pst-all,pstricks,psfig,fancybox, fullpage,epsf,epsfig}
\makeatletter

\def\@abssec#1{\vspace{.05in} \parindent 0in {\bf #1. }\ignorespaces}
\def\abstract{\@abssec{Abstract}}

\def\keywords{\@abssec{Key words}}

\def\AMSMOS{\@abssec{AMS(MOS) subject classifications}}

\def\@begintheorem#1#2{\par\vspace{.07in}\par\bgroup{\sc\bf #1\ #2. }\it\ignorespaces}
\def\@opargbegintheorem#1#2#3{\par\bgroup{\sc #1\ #2\ (#3). }\it\ignorespaces}
\def\@endtheorem{\egroup\vspace{.07in}}

\newtheorem{definition}{Definition}

\makeatother

\def\nowidth#1{{\setbox0=\hbox{#1}\hspace*{-\wd0}\hbox to 0pt {#1}}}

\def\cliquev{
\begin{pspicture}(-1,-2)(1,2)
\multiput(0,2)(0,-1){4}{\psdot[linewidth=1.5pt](0,0)}
{\psset{linewidth=.1pt}
\multiput(0,2)(0,-1){3}{\psbezier(0,0)(.2,-.5)(0,-1)}
\multiput(0,2)(0,-1){2}{\psbezier(0,0)(.5,-1)(0,-2)}
\multiput(0,2)(0,-1){1}{\psbezier(0,0)(.8,-1.5)(0,-3)}
}\end{pspicture}
}

\def\cliqueh{
\begin{pspicture}(-3,-1)(3,1)
\multiput(-3,0)(1,0){5}{\psdot(0,0)}
{\psset{linewidth=.1pt}
\multiput(-3,0)(1,0){4}{\psbezier(0,0)(.5,.2)(1,0)}
\multiput(-3,0)(1,0){3}{\psbezier(0,0)(1,.5)(2,0)}
\multiput(-3,0)(1,0){2}{\psbezier(0,0)(1.5,.8)(3,0)}
\multiput(-3,0)(1,0){1}{\psbezier(0,0)(2,1.1)(4,0)}
}
\end{pspicture}
}

\def\graph{
\begin{pspicture}(0,-1.5)(6,1.5)
\rput(1,-.5){\cliquev}\rput(2,-.5){\cliquev}\rput(3,-.5){\cliquev}\rput(4,-.5){\cliquev}\rput(5,-.5){\cliquev}
\rput(4,-1.5){\cliqueh}\rput(4,-.5){\cliqueh}\rput(4,.5){\cliqueh}\rput(4,1.5){\cliqueh}
\rput[tr](.88,1.4){$0$}\rput[tr](.9,.4){$1$}\rput[tr](.9,-1.6){$n_1-1$}
\cput*[framesep=0pt](.7,-.5){$i_1$}
\rput[tr](1.9,1.4){$1$}\rput[tl](5.2,1.4){$n_2-1$}
\cput*[framesep=0pt](3,1.8){$i_2$}
\cput*[framesep=1pt](3,-.2){\large $x$}
\end{pspicture}
}

\begin{document}
\title{Sharp Bounds for Bandwidth of Clique Products} %!PN

\author{Tanya Y. Berger-Wolf\thanks{Dept. of Comp. Sci.,
                               Univ. of Illinois at Urbana-Champaign,
                               1304 W. Springfield Avenue,
                               Urbana, Illinois 61801, USA.
			Supported in part by an NSF Graduate Fellowship.
                             {\tt tanyabw@cs.uiuc.edu}}
%\and L. H. Harper\thanks{Dept. of Math.,
%                               Univ. of California,
%			       Riverside, CA 92521, USA.
%   				{\tt harper@math.ucr.edu}}
\and Mitchell A. Harris\thanks{Dept. of Comp. Sci.,
                               Univ. of Illinois at Urbana-Champaign,
                               1304 W. Springfield Avenue,
                               Urbana, Illinois 61801, USA.
   				{\tt maharri@cs.uiuc.edu}}}
\maketitle

\pagestyle{myheadings}
\markboth{}{} 

\begin{abstract} \small\baselineskip=10pt
The bandwidth of a graph is the labeling of vertices with minimum
maximum edge difference. For many graph families this is
NP-complete. A classic result computes the bandwidth for the
hypercube. We generalize this result to give sharp lower bounds for
products of cliques. This problem turns out to be equivalent to one in
communication over multiple channels in which channels can fail and
the information sent over those channels is lost.  The goal is to
create an encoding that minimizes the difference between the received
and the original information while having as little redundancy as
possible.  Berger-Wolf and Reingold \cite{BR} have considered the
problem for the equal size cliques (or equal capacity channels).  This
paper presents a tight lower bound and an algorithm for constructing
the labeling for the product of any number of arbitrary size cliques.
\end{abstract}


%\begin{AMSMOS}
%94A05, 94A24, 94A34, 68Q25, 68R10
%\end{AMSMOS}

%-------------------------Introduction--------------------------

\section{Introduction}
Labeling of graph vertices is an active area of research related to
many applications ranging from VLSI to computational biology. There
are several graph parameters associated with a labeling that can be
optimized. One such is bandwidth, the maximum difference between
labels on an edge. In general, bandwidth of a graph is an NP-complete
problem \cite{papa}. Even for very restricted families, e.g. trees of
maximum degree 3 or varieties of caterpillars, it remains
NP-complete. In this paper we focus on the bandwidth of Hamming graphs
-- cartesian product of cliques. Applications of this specific
problems arise in designing encodings for packet-switched networks
that minimize the error in case of packet loss \cite{batllo,jayant1,jayant2,yang}.

\subsection{Problem Statement\\}
Given a graph, $G=(V,E)$, a {\em labeling} $f$ of a graph is an assignment
of numbers $\{1,...,|V|\}$ to the graph's vertices:
\[ f:V \to \{1,...,|V|\} \]
A labeling $f$ is a bijection.

Given a labeling, {\em bandwidth} is the maximum over all edges of the
difference between labels on an edge:
\[ B_f(G) = \max_{\langle u,v\rangle\in E}{\{|f(u) - f(v)|\}} \]
{\em Graph bandwidth} is the minimum possible bandwidth of a graph:
\[ B(G) = \min_f{\{B_f(G)\}} \]
The Bandwidth Optimization problem is the problem of finding a
labeling that minimizes the graph bandwidth. As we have mentioned, for
a general graph, the bandwidth optimization problem is NP-hard
\cite{papa}.

A {\em cartesian product} of two graphs $G_1 = (V_1,E_1)$ and
$G_2 = (V_2,E_2)$ is a graph whose vertices are tuples of the original
vertices, and whose edges go between vertex tuples different in only
one coordinate:
\[ G_1\times G_2 = \left({\begin{array}{c}
		V = \{(v_1,v_2) : v_1\in V_1, v_2\in V_2\},\\
		E = \{\langle (v_1,v_2), (u_1,u_2)\rangle :
			v_1 = u_1 \wedge \langle v_2,u_2\rangle\in E_2
		\mbox{ or }\langle v_1,u_1\rangle\in E_1\wedge v_2=u_2\}
			  \end{array}}\right)
\]
A cartesian product can be inductively extended to more than two graphs.

A {\em clique} $K_n$ is a simple undirected graph on $n$ vertices with
${n\choose 2}$ edges, an edge between every vertex pair.  Given $d$
complete graphs (cliques) $K_{n_1},...,K_{n_d}$, a {\em Hamming graph}
is their cartesian product $K_{n_1}\times\cdots\times K_{n_d}$.

We first consider the product of two cliques of unequal order. We
prove a tight lower bound on the graph bandwidth and give an optimal
algorithm that achieves that lower bound. We generalize the results
for arbitrary number of cliques. To the best of our knowledge, this
is the first result for bandwidth optimization of products of cliques
of unequal order.

\subsection{Problem Background\\}
In graph theory, the bandwidth problem was introduced by Harper in
1966 \cite{harper2}, where he solved the problem for hypercubes, that
is products of $K_2$'s. Hendrich and Stiebitz \cite{HS} solved the
bandwidth problem for products of two cliques of equal size. In
\cite{harper3} Harper gives a non-constructive asymptotically best
lower bound for products of cliques of equal sizes. Berger-Wolf and
Reingold \cite{BR} have introduced a general technique that gives a
lower bound and an algorithm for $d$-fold products of cliques of equal
sizes. While their technique is applicable to cliques of unequal
sizes, the lower bound is very loose in that case. Here we propose a
new and simple technique for deriving a tight lower bound and give an
optimal algorithm for the the case of unequal size cliques.

%--------------------------------Results-------------------------

\section{Results}
We present a technique that provides a lower bound for the bandwidth
of the Hamming graph as a maximum of lower bounds for each clique. The
technique also suggests an algorithm which provides an almost matching
upper bound and is thus nearly optimal. The minimal bandwidth is
\[B(K_{n_1}\times K_{n_2}\times\dots\times K_{n_d})=\Theta(B(K_2^d)\times\prod{n_t\over2}),\]
where $B(K_2^d)=\sum_{t=0}^{d-1}{t\choose \lfloor{t/2}\rfloor}$ is the
bandwidth of the product of $d$ $2$-cliques.

The problem of minimizing the bandwidth of $K_{n_1}\times
K_{n_2}\times\cdots\times K_{n_d}$ can be thought of as the problem of
arrangement of numbers $\{1,\dots, \prod{n_t}\}$ in an $n_1\times
n_2\times\cdots\times n_d$ matrix in a way that minimizes the maximum
difference between the largest and the smallest number in any {\em
line} -- a full one-dimensional submatrix. The correspondence is
straightforward; the numbers within a line represent vertices within
the same clique and so a minimizing arrangement minimizes the
bandwidth.

\begin{figure}
\begin{center}
\begin{pspicture}(-2,-1.5)(10,1.5)
\begin{pspicture}(1,1.2)(3,3)
\multiput(0,0)(0,-1){4}{\multiput(1,2.5)(1,0){5}{\psdot(0,0)}}
\psframe(.8,-.7)(5.2,2.7)
\rput[r](.7,2.5){$1$}\rput[r](.7,1.5){$2$}\rput[r](.7,-.5){$n_1$}
\cput*[framesep=0pt](.5,.5){$i_1$}
\rput[b](1,2.85){$1$}\rput[b](2,2.85){$2$}\rput[b](5,2.85){$n_2$}
\rput(3,3){$i_2$}
\cput*[framesep=0pt](3,.5){\large $x$}
\end{pspicture}
\psline[doubleline=true]{<->}(2.5,-.5)(4.5,-.5)
\rput(7,-.2){\graph}
\end{pspicture}
\end{center}
\caption{\small Correspondence between the number arrangement of a two-dimensional matrix and the bandwidth minimization of a product of two cliques problems.}
\label{correspondence}
\end{figure}
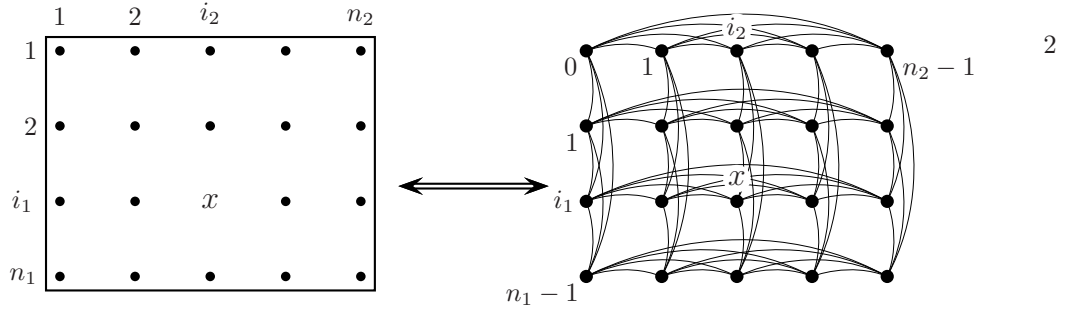

Figure \ref{correspondence} shows the correspondence between the two
problems in case of two dimensions. We assume throughout this paper
without loss of generality that $n_1 \le n_2 \le \cdots \le n_d$.

We first show a lower bound for the problem and then present an
algorithm that nearly achieves that lower bound. After giving the
fundamental lemmata we demonstrate the approach for the
two-dimensional case and the generalize it to arbitrary dimensions.

\begin{definition}
Let an {\em arrangement} be a one-to-one function
$A: {\mathcal N} \to {\mathcal N}^d$ 
from $\{1, \dots, \prod_{t=1}^d{n_t}\}$ onto the set
of cells of an $n_1\times n_2\times\cdots\times n_d$ matrix.
Let a {\em line} be a full one-dimensional submatrix of the
type $(i_1,i_2,...,*,...,i_d)$ with all but one coordinate fixed.
Then the {\em spread} of an arrangement is the maximum difference
over all lines between any two numbers in any line:
\[{\rm spread}(A) = 
\max_{1\le i_t,j_t\le n_t, 1\le t\le d \atop s\neq t \rightarrow i_s=j_s}
{|A^{-1}(i_1,i_2,...,i_t,...,i_d) - A^{-1}(j_1,j_2,...,j_t,...,j_d)|}.\]
\end{definition}

Since $A$ is a bijection, to simplify the notation, we will use $A$
and $A^{-1}$ interchangeably, the meaning hopefully being clear
from the context.

First, we note that the lower bound on the spread for
{\em any} line is the lower bound on the spread in the entire matrix,
therefore the maximum of the line bounds is also a lower bound for the
matrix spread. Thus we can deal with one line at a time. We then
restrict our attention to a special kind of arrangement showing that
this restriction does not eliminate optimal arrangements. Then for
these arrangements it is easier to find a line with a large spread.

\begin{definition}
\label{monotonic_definition}
An arrangement is {\em monotonic} if the values in any line ascend
with the increase of the changing coordinate. That is, an arrangement
$A$ is monotonic if for all $1\le i_t,j_t\le n_t, \: 1\le t\le d$
\[((s\neq t\rightarrow i_s=j_s) \wedge i_t < j_t)\rightarrow
A(i_1,...,i_t,...,i_d) < A(j_1,...,j_t,...,j_d).\]
\end{definition}

%-----------------------Monotonic Lemma-----------------------------

\begin{lemma}
\label{monotonic_lemma}
Given any arrangement of any set of $n_1n_2\cdots n_d$ numbers,
sorting it to become monotonic one coordinate at a time, one line at a
time, does not increase the spread. That is, for any arrangement $A$,
\[{\rm spread}({\rm sorted}(A)) \le {\rm spread}(A).\]
\end{lemma}
\begin{proof}
We first show that given any arrangement, sorting the numbers to
become monotonic in {\em one} coordinate does not increase the overall
spread. It is obvious that rearranging the numbers in any way within
the same line does not change the spread in that line, thus sorting
within a coordinate does not change the spread in that coordinate.
Suppose the spread has increased in another coordinate. The situation
is illustrated in Figure \ref{rearrange_monotonic}.  Let the maximum
spread in that coordinate {\em after} sorting be $b_t - a_s$ appearing in
line $j$ (where $b_t$ was in line $t$ {\em before} the rearrangement,
and $a_s$ was in line $s$). Then
\[|b_t - a_t| < b_t - a_s, \mbox{ thus }a_s < a_t, \mbox{ and}\]
\[|b_s - a_s| < b_t - a_s, \mbox{ thus }b_s < b_t.\]
Then there are $j-2$ (since $b_t$ and $a_s$ are now in line $j$) $b$'s
less than $b_t$ and not equal to $b_s$. There are $j-1$ $a$'s less
than $a_s$ and not equal to $a_t$. Therefore, by the pigeonhole
principle, there exists $a_p < a_s$ that was paired up with $b_p > b_t$
before the rearrangement. But then $b_p - a_p > b_t - a_s$, which
contradicts the assumption that the spread increased after
sorting. Thus sorting in one coordinate does not increase the spread
in {\em any} coordinate.
\begin{figure}
\begin{center}
{\psset{unit=.3}
\begin{pspicture}(0,0)(12,12)
\psframe(0,0)(12,12)
\psframe(0,3)(12,4)\psframe(0,8)(12,9)
{\psset{fillstyle=solid,fillcolor=lightgray}
\psframe(0,8)(5,9)\psframe(0,3)(2,4)\psframe(3,3)(5,4)
}
\psframe(2,0)(3,12)\psframe(5,0)(6,12)\psframe(9,0)(10,12)
\rput(2.5,12.5){$s$}\rput(5.5,12.5){$j$}\rput(9.5,12.5){$t$}
\rput(2.5,8.5){$a_s$}\rput(5.5,8.5){$a_s$}\rput(9.5,8.5){$a_t$}
\pscurve{->}(2.5,9)(4,9.5)(5.5,9)
\pscurve{->}(9.5,9)(9.25,9.75)(8.5,9.5)\pscurve{->}(9.5,9)(9.75,9.75)(10.5,9.5)
\rput(2.5,3.5){$b_s$}\rput(5.5,3.5){$b_t$}\rput(9.5,3.5){$b_t$}
\pscurve{<-}(5.5,4)(7.5,4.5)(9.5,4)
\pscurve{->}(2.5,4)(2.25,4.75)(1.5,4.5)\pscurve{->}(2.5,4)(2.75,4.75)(3.5,4.5)
\end{pspicture}
}
\end{center}
\caption{Sorting the values within the rows causes the spread in columns
to increase to $b_t - a_s$, occurring now in column $j$. Before
sorting, $b_t$ was in column $t$ and $a_s$ was in column $s$. Shaded
are the $c$'s less than $a_s$ but not equal to $a_t$ and $b$'s that
are less than $b_t$ and not equal to $b_s$. Note that necessarily $s <
t$, but $j$ can be any column relative to $s$ and $t$.}
\label{rearrange_monotonic}
\end{figure}
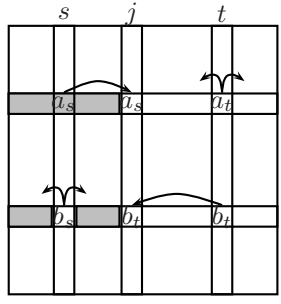

Gale and Karp \cite{GK} show that if the arrangement was monotonic in
any coordinate then it will remain so after the numbers are sorted in
any other coordinate. Thus the matrix can be sorted to have a
monotonic arrangement one coordinate at a time, one line at a time,
without increasing the spread.
\end{proof}

This allows us to restrict attention to monotonic arrangements. From
these arrangements we can more easily find a general structure of the
lower bound on the spread.

Consider some number $x$ in a cell of a monotonic arrangement. The $d$
axis-parallel hyperplanes that pass through that cell divide the
matrix into $2^d$ orthants. For any monotonic arrangement of any set
of $n_1n_2\cdots n_d$ numbers, all the numbers in the orthant
containing the coordinate $(1,1,...,1)$ are necessarily less than $x$,
and all the numbers in the orthant containing the last coordinate are
necessarily greater than $x$.  Besides the first and the last
orthants, the other $2^d-2$ orthants may contain both numbers less
than and greater than $x$. Any line passing through $x$ necessarily
has both numbers less and greater than $x$ by the nature of
monotonicity of the arrangement. However any other line can be filled
entirely with only smaller or larger numbers.

%-----------------Lower bound Lemma------------------------------

\begin{lemma}
\label{monotonic_lower}
For the optimal arrangement $A$ of numbers
$1,...,\prod_{t=1}^{d}{n_t}$ in an $n_1\times n_2\times ...\times n_d$
matrix, there exists a line $(i_1,i_2,...,i_{j-1},*,i_{j+1},...,i_d)$ (all the
coordinates but the $j$th are fixed) in that arrangement and there
exists a cell in that line $(i_1,i_2,...,i_j,...,i_d)$ such that the
spread in that line is at least the volume of any minimal set of
orthants (as defined by the cell) that separates between the orthant
containing the cell $(1,..,1)$ (the first orthant) and the orthant
containing the cell $(n_1,...,n_d)$ (the last orthant).\\
%\noindent (*** the orthants have to be defined more precisely - where do the dividing hyperplanes belong *** tbw)

\end{lemma}
\begin{proof}
First, we will note several facts: 
\begin{itemize}
\item Removing any {\em minimal} separating set of orthants leaves 
only two {\em connected} sets of orthants: the set containing the
first orthant (we shall call this set ``small'' orthants) and the set
containing the last orthant (``large'' orthants).
%(*** Should this be proved or is it obvious? *** tbw).

\item Since the set is a minimal separating set, {\em any} cell within any
of the separating orthants is contained in lines that intersect the
``large'' orthants and in lines that intersect the ``small'' orthants.

\item No line passes through both the ``small'' and ``large'' orthants,
since otherwise they would not be separated.
\end{itemize}

Now we are ready to prove the lemma. Let $V_{small}(cell)$ be the
volume (number of cells) of the ``small'' orthants, $V_{large}(cell)$
be the volume of the ``large'' orthants, and $V_{sep}(cell)$ be the
volume of the separating orthants. Note that
$V_{small} + V_{sep} + V_{large} = \prod_{t=1}^{d}{n_t} = V$. 
For the optimal arrangement $A$ let $l=(i_1,i_2,...,*,...,i_d)$ be 
the line with the largest spread (that is, the spread of the
arrangement is the spread in this line). Here are the two possible cases:
\begin{itemize}
\item there exists a cell $(i_1,i_2,...,i_j,...,i_d)$ such that the
smallest number in the line, $\min_l$, is at most $V_{small}(cell)$
(for any separating set defined by the cell), and the largest number
in the line, $\max_l$, is at least $V-V_{large}(cell)$. Then
\begin{eqnarray*}
{\rm spread}(A) &\ge& {\rm max}_l - {\rm min}_l\\
&\ge& V-V_{large}(cell)-V_{small}(cell)\\
&=& V_{sep}(cell)
\end{eqnarray*}
and the statement of the lemma holds.

\item for all cells in the line, for some separating set for each cell,
either $\max_l < V-V_{large}$ or $\min_l > V_{small}$.\\

Let $i_j$ be such that $V_{sep}$ as defined by $(i_1,...,i_j,...,i_d)$
is the smallest.  Without loss of generality we assume that $\min_l >
V_{small}(i_1,...,i_j,...,i_d)$, while  $\max_l$ can be either less or
greater or equal to $V-V_{large}(i_1,...,i_j,...,i_d)$.

Since $\min_l > V_{small}(i_1,...,i_j,...,i_d)$ there must be at least
one element less than $\min_l$ in the separating orthants defined by
the cell $(i_1,...,i_j,...,i_d)$. Suppose there are $s$ elements that
are less than $\min_l$ total in the separating orthants. Then there
are at most $s-1$ elements greater than $\min_l$ in the small
orthants. Each of the $s$ elements in the separating orthants must be
in a line that intersects large orthants (as defined by the cell
$(i_1,...,i_j,...,i_d)$). Since $\max_l - \min_l$ is the largest
spread, all elements in those lines must be less than $\max_l$. Let
there be $l$ of those elements. One possible way this can happen is
shown in Figure \ref{v_sep}.
\begin{figure}[t]
\begin{center}
\epsfig{file=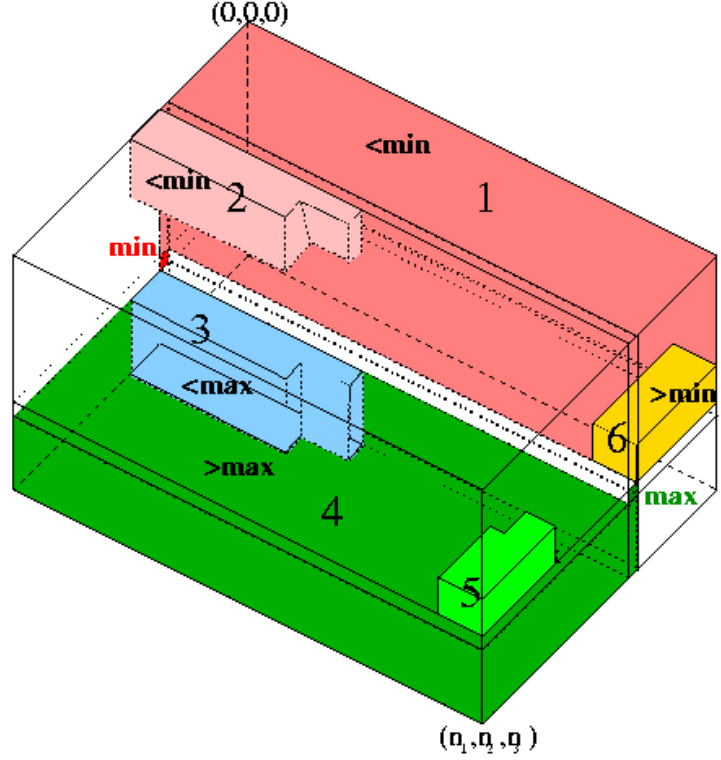, height=4.1in}
\end{center}
\caption[$max_l - min_l < V_{sep}$.]{%\small
Areas (1) and (6) are the small orthants, areas (4) and (3) are the
large orthants, and the uncolored area, with (2) and (5), are the
separating orthants. The indicated min and max are the minimum and
maximum in the line $l$.  (1) are the numbers less than $\min_l$ in
the small orthants.  (2) are the $s$ numbers less than $\min_l$ in the
separating orthants. (3) are the $l$ numbers less than $\max_l$ in the
large orthants. Those are the intersection of the lines that have a
light red minimum with the large orthants. (4) are the numbers greater
than $\max_l$ in the large orthants. (5) are the $p$ numbers greater
than $\max_l$ in the separating orthants.  (6) are the matching
numbers greater than $\min_l$ in the small orthants.}
\label{v_sep}
\end{figure}
We use a switching idea similar to Fishburn, Tetali, and Winkler
\cite{FTW}. Suppose there are $t \ge l$ elements total less than
$max_l$ in the large orthants. Then $max_l \le V-V_{large}+t$. Replace
the largest of those $t$ elements $l_1$ with $max_l$, then replace the
next largest element $l_2$ with $l_1$ and so on, trickling down until
we get to the smallest of those $t$ elements. Put that smallest
element instead of $max_l$. We have not violated the monotonicity. We
have increased each of the $t$ elements by at least $1$ and decreased
$max_l$ by $t$.  Therefore the new $max_l \le V-V_{large}$ and all the
elements in the large orthants are greater than $max_l$. Similarly, we
can replace the largest of the $s$ small elements, $s_1$, with
$min_l$, then replace the next largest element $s_2$ with $s_1$, and
so on until we either reach the $t$th largest or the smallest of the
$s$ elements. We replace $min_l$ with that element. We have increased
each of the $s$ elements by at most $1$, so the relative spread has
not increased. The minimum $min_l$ has decreased by at most $t$, so
the spread in line $l$ has not increased.

If $s > t$ then we have stopped after replacing $t$ of the $s$
elements and there are still some elements less than the new $min_l$
in the separating orthants in the lines with elements greater than the
new $max_l$.  We have not increased the spread in line $l$ or anywhere
else, but the spread in those lines is greater than the new $max_l -
min_l$ which equals the old spread since both the minimum and the
maximum decreased by the same amount. This is a contradiction to the
assumption that $l$ was the line with the maximum spread.

If $s \le t$ then there are no elements less than $min_l$ in the
separating orthants and the new $min_l \le V_{small}$. If there are no
elements greater than $max_l$ in the separating orthants, then $max_l
- min_l \ge V_{sep}$, which means the initial spread was also at least
$V_{sep}$, which is a contradiction. Suppose there are $p$ elements
greater than the new $max_l$ in the separating orthants. Those
elements must be in lines that intersect small orthants and the
elements in the small orthants in those lines must be greater than the
new $min_l$. Suppose there are $q$ of those elements. We can perform
the same replacement procedure and if $p > q$ we will get the same
contradiction as in case of $s > t$. Otherwise, $p \le q$. Since there
are no elements less than the new $min_l$ in the separating orthants
and there are $q$ elements in the small orthants that are greater than
$min_l$, then the new $min_l = V_{small} - q$ and therefore the
original $min_l = V_{small} - q + s$. Similarly, since there are $p$
elements greater than the new $max_l$ in the separating orthants and
no elements less than it in the large orthants, the new $max_l = V -
V_{large} - p$ and thus the original $max_l = V - V_{large} - p +
t$. So the original spread is the difference between the original
$\max_l$ and $\min_l$ which is
\[ (V - V_{large} - p + t) - (V_{small} - q + s) = V_{sep} + q-p + s-t.
\]
Since $p \le q$ and $s \le t$, the original spread was at least
$V_{sep}$, which is, again, a contradiction.
\end{itemize}

Therefore, the spread in an optimal arrangement is at least $V_{sep}$
for any minimal set of the separating orthants as defined by some cell
in the maximum spread line.

\end{proof}

\begin{figure}
\begin{center}
\begin{pspicture}(-5,-2)(5,3)
\rput(-7,-2){\psset{unit=.5}
\psline[linestyle=dashed](2.5,7.5)(2.5,2.5)(9.5,2.5)
\psline[linestyle=dashed](0,0)(2.5,2.5)
{\psset{linewidth=2pt}
\psframe(0,0)(7,5)
\psline(0,5)(2.5,7.5)(9.5,7.5)(9.5,2.5)(7,0)\psline(7,5)(9.5,7.5)
}
\rput(4.5,3.75){\large$\mathbf x$}
\psset{unit=.7}\rput(2,2){
\psset{linewidth=.5pt}
\psline[linestyle=dashed](2.5,7.5)(2.5,2.5)(9.5,2.5)
\psline[linestyle=dashed](0,0)(2.5,2.5)
\psframe(0,0)(7,5)
\psline(0,5)(2.5,7.5)(9.5,7.5)(9.5,2.5)(7,0)\psline(7,5)(9.5,7.5)
\psset{linewidth=3pt}
\psdot[dotstyle=o](2.5,7.5)
\psdots[dotstyle=*](0,0)(7,5)(9.5,2.5)(7,0)
\psdots[dotstyle=asterisk](0,5)(2.5,2.5)(9.5,7.5)
}}
\rput(4,.5){\psset{unit=.5}
\psset{linewidth=.8pt}
\psline(2,-1)(2,1)(0,3)(-2,1)(-2,-1)(0,-3)(2,-1)(0,1)(-2,-1)
\psline(-2,1)(0,-1)(2,1)\psline(0,-1)(0,-3)\psline(0,1)(0,3)
\psset{linewidth=3pt}
\psdot[dotstyle=o](0,3)
\psdots[dotstyle=*](-2,-1)(0,-1)(2,-1)(0,-3)
\psdots[dotstyle=asterisk](-2,1)(0,1)(2,1)
\rput[l](2.5,3){SMALL}\rput[l](2.5,-2){LARGE}
}
\end{pspicture}
\end{center}
\caption{Correspondence between the orthants and a hypercube in $3$
dimensions. The $small(x)$ and $large(x)$ numbers meet only if a line
passes through two adjacent orthants.}
\label{orthants_cube}
\end{figure}
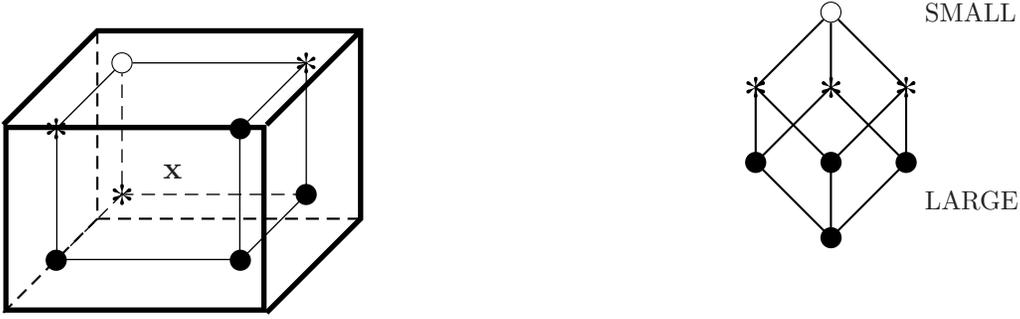

We have proved that the spread in an optimal arrangement is at least
the volume of {\em any} minimal set of orthants separating between the
``small'' and ``large'' orthants, for some cell in the largest spread
line. Therefore, there is a cell such that the spread is at least
the volume of the {\em largest} minimal separating set of orthants,
the one that has the largest number of orthants. In fact, if we associate a
super vertex with each orthant and have an edge between any two
vertices if the corresponding orthants are adjacent, then we get a
$d$-dimensional hypercube representing the orthants. 
Figure \ref{orthants_cube} shows this in $3$ dimensions.
By definition of bandwidth, the largest minimum separating set of
orthants is exactly the bandwidth of the $d$-dimensional hypercube. So
for an optimal arrangement, for any line in the arrangement, the
spread is at least the minimum over all cells of the volume of the
bandwidth-separating set of orthants. Thus the spread in the optimal
arrangement is at least
\[\max_{\mbox{\footnotesize all lines}}{\{
    \min_{\mbox{\footnotesize cells in a line}}{\{
      \min_{\mbox{\footnotesize separating set }\atop
	    \mbox{\footnotesize of }B(K_2^d)\mbox{ \footnotesize orthants}}{\{
        \mbox{volume of the separating set of orthants}
      \}}
    \}}
  \}}.\]

Using this lemma we can calculate the lower bound on the spread in the
optimal arrangement. We will first demonstrate our approach in two
dimensions and then generalize is to arbitrary number of dimensions.

%--------------------------------2D-----------------------------

\subsection{Two Dimensions\\}
\begin{theorem}
\label{two_dim_lower}
Without loss of generality assume $n_1\le n_2$. The spread in any
arrangement of an $n_1$ by $n_2$ matrix is at least
\[{(n_1+1)n_2\over 2} - 1 \mbox{ if }n_1\mbox{ is odd, }\]
\[{n_1(n_2+1)\over 2} - 1 \mbox{ if }n_1\mbox{ is even}.\]
\end{theorem}

\begin{proof}
In two dimensions, there is only one set of orthants separating
between the first and the last orthants, which is the other two of the
four orthants. This is also consistent with $B(K_2^2)=2$. Thus the
spread in a two dimensional arrangement is at least
\begin{eqnarray*}
\lefteqn{\max_{\mbox{\footnotesize all lines}}{\{
    \min_{\mbox{\footnotesize cells in a line}}{\{
        \mbox{area of the two separating orthants}
    \}}
  \}} = }\\\\
& &  \max{\left\{
	\begin{array}{r}
          {\displaystyle \max_{\mbox{\footnotesize all rows}}{\{
            \min_{\mbox{\footnotesize cells in a row}}{\{
              \mbox{area of the two separating orthants}
            \}}\}},} \\
          {\displaystyle \max_{\mbox{\footnotesize all columns}}{\{
            \min_{\mbox{\footnotesize cells in a column}}{\{
              \mbox{area of the two separating orthants}
            \}}\}}}
	\end{array}
     \right\}}.
\end{eqnarray*}

\begin{figure}[t]
\begin{minipage}{7cm}
\begin{center}
\begin{pspicture}(-1,-1.5)(7,3)
\psframe(.8,-.7)(5.2,2.7)
\rput[r](.7,2.5){$1$}\rput[r](.7,-.5){$n_1$}
\rput[b](1,2.85){$1$}\rput[b](5,2.85){$n_2$}
{\psset{fillstyle=solid}
\psframe[fillcolor=lightgray](.8,.35)(3.15,2.7)\rput(1.95,1.55){\psframebox*{SMALL}}
\psframe[fillcolor=darkgray](2.85,.65)(5.2,-.7)\rput(4.05,-.05){\psframebox*{LARGE}}
}
\rput(3,.5){\psframebox*[framesep=1.5pt]{\large $x$}}
\rput(1.95,-.05){??}\rput(4.05,1.55){??}
\end{pspicture}
\end{center}
\caption{The location of numbers less and greater than a given cell value in a monotonic arrangement.}
\label{monotonic}
\end{minipage}
\hfill
\begin{minipage}{7cm}
\begin{center}
\begin{pspicture}(-1,-1)(7,3)
\psframe(.8,-.7)(5.2,2.7)
\rput[r](.7,2.5){$1$}\rput[r](.7,.5){$i_1$}\rput[r](.7,-.5){$n_1$}
\rput[b](1,2.85){$1$}\rput[b](3,2.85){$i_2$}\rput[b](5,2.85){$n_2$}
{\psset{fillstyle=solid,dimen=outer}
\psframe[fillcolor=lightgray](.8,.65)(2.85,2.7)
\rput(1.8,1.55){\psframebox*[framesep=1pt]{\parbox[c]{1.4cm}{\center\scriptsize$(i_1-1)\times (i_2-1)$}}}
\psframe[fillcolor=darkgray](3.15,.35)(5.2,-.7)
\rput(4.15,-.02){\psframebox*[framesep=1pt]{\parbox[c]{1.5cm}{\center\scriptsize$(n_1-i_2)\times(n_2-i_2)$}}}
}
\psframe[fillcolor=lightgray,fillstyle=hlines](2.85,2.7)(3.15,.65)
\psline(.8,.65)(5.2,.65)\psline(.8,.35)(5.2,.35)\psline(2.85,-.7)(2.85,2.7)\psline(3.15,-.7)(3.15,2.7)
\psdot(1,.5)\psdot(5,.5)
%\rput(3,.5){\large $x$}
\rput(4,2.4){$\underbrace{\hskip 2.3cm}_{n_2-i_2+1}$}
\rput(5.1,1.55){$\scriptstyle{i_1}\left\{\begin{array}{l}\\\\\\\\\\\end{array}\right.$}
\rput(1.95,-.4){$\overbrace{\hskip 2.3cm}^{i_2}$}
\rput[l](.4,-.05){$\left.\begin{array}{l}\\\\\\\end{array}\right\}\scriptstyle{n_1-i_1+1}$}
%\rput(1.8,.2){\tiny$(n_1-i_2-1)\times i_2$}
\end{pspicture}
\end{center}
\caption{Spread in a row as the volume of the unfilled separating orthants
(as defined by the cell $(i_1,i_2)$). To minimize the spread, the larger of
the top and bottom parts of the column is filled.}
\label{spread2}
\end{minipage}
\end{figure}

Let $1\le i_1\le n_1$, $1\le i_2\le n_2$. The lower bound on the
spread in an $n_1$ by $n_2$ matrix is (see Figure \ref{spread2} for
illustration of the calculations)
\begin{eqnarray*}
\max{\left\{{
\begin{array}{l}
{\displaystyle \max_{\mbox{\scriptsize row }i_1}{\left\{{
	\min_{i_2}{\left\{{
		i_1(n_2-i_2+1)+(n_1-i_1+1)i_2-2-\max{\{i_1-1,\; n_1-i_1\}}
						}\right\}}
					}\right\}}},\\\\
{\displaystyle \max_{\mbox{\scriptsize col }i_2}{\left\{{
	\min_{i_1}{\left\{{
		i_1(n_2-i_2+1)+(n_1-i_1+1)i_2-2-\max{\{i_2-1,\; n_2-i_2\}}
						}\right\}}
					}\right\}}}
\end{array}
}\right\}}
\end{eqnarray*}
The spread in a line is a symmetric unimodal function of the free
coordinate, with the maximum occurring in the middle, thus we separate
at the half point and evaluate at endpoints:
\begin{eqnarray*}
&=&\max{\left\{{
\begin{array}{l}
{\displaystyle \max_{\mbox{\scriptsize row }i_1}{\left\{{
	\begin{array}{lcl}
	i_1\le\left\lceil{n_1\over2}\right\rceil & {\stackrel{\Longrightarrow}{\scriptstyle i_2=1}} &
		i_1n_2+(n_1-i_1+1)-(n_1-i_1)-2\\\\
	i_1 > \left\lceil{n_1\over2}\right\rceil & {\stackrel{\Longrightarrow}{\scriptstyle i_2=n_2}} &
		i_1+(n_1-i_1+1)n_2-(i_1-1)-2
	\end{array}
					}\right\}},}\\\\
{\displaystyle \max_{\mbox{\scriptsize col }i_2}{\left\{{
	\begin{array}{lcl}
	i_2\le\left\lceil{n_2\over2}\right\rceil & {\stackrel{\Longrightarrow}{\scriptstyle i_1=1}} &
		(n_2-i_2+1)+n_1i_2-(n_2-i_2)-2\\\\
	i_2 > \left\lceil{n_2\over2}\right\rceil &{\stackrel{\Longrightarrow}{\scriptstyle i_1=n_1}} &
		n_1(n_2-i_2+1)+i_2-(i_2-1)-2
	\end{array}
					}\right\}}}
\end{array}
	}\right\}}\\\\\\
&=&\max{\left\{{
\max_{\mbox{\scriptsize row }i_1}{\left\{{
	i_1\le\left\lceil{n_1\over2}\right\rceil \:\Rightarrow\:
		i_1n_2-1  \hfill\atop
	i_1 > \left\lceil{n_1\over2}\right\rceil \:\Rightarrow\:
		(n_1-i_1-1)n_2-1
					}\right\}}, \;\;
\max_{\mbox{\scriptsize col }i_2}{\left\{{
	i_2\le\left\lceil{n_2\over2}\right\rceil \:\Rightarrow\:
		n_1i_2-1  \hfill\atop
	i_2 > \left\lceil{n_2\over2}\right\rceil \:\Rightarrow\:
		n_1(n_2-i_2-1)-1
					}\right\}}
	}\right\}}\\\\\\
&=&\max{\left\{{
\max{\left\{\left\lceil{n_1\over2}\right\rceil n_2-1, \:
	\left\lfloor{n_1\over2}\right\rfloor n_2-1\right\}}, \:\:
\max{\left\{n_1 \left\lceil{n_2\over2}\right\rceil - 1, \;\;
	n_1\left\lfloor{n_2\over2}\right\rfloor - 1\right\}}
	}\right\}}\\\\\\
&=&\max{\left\{{
\left\lceil{n_1\over2}\right\rceil n_2 - 1,\: n_1
\left\lceil{n_2\over2}\right\rceil - 1 }\right\}}\\\\\\
&=&\left\{{
\begin{array}{l}
	n_1\mbox{ even }\:\Rightarrow\:
	{n_1n_2 \over2}-1 \le n_1 \left\lceil{n_2\over2}\right\rceil - 1
		\:\Rightarrow\: {
			\begin{array}{l}
			n_2\mbox{ \small odd }\:\Rightarrow\:{n_1(n_2+1)\over2}-1\\
			n_2\mbox{ \small even }\:\Rightarrow\:{n_1n_2\over 2}-1
			\end{array}
			}\\\\
	n_1\mbox{ odd }\:\Rightarrow \:
	{(n_1+1)n_2\over2}-1 \ge n_1 \left\lceil{n_2\over2}\right\rceil - 1
		\:\Rightarrow\: {(n_1+1)n_2\over2}-1
\end{array}
}\right\}%\\
\end{eqnarray*}

Thus in case of $n_1$ odd the spread in the matrix is at least
$(n_1+1)n_2/2-1$ and if $n_1$ is even and $n_2$ is odd then the spread
is at least $n_1(n_2+1)/2-1$. We will present arrangements that
achieve these bounds thus the lower bound is sharp. We now show,
however, that the lower bound of $n_1n_2/2-1$ for the case of both
$n_1$ and $n_2$ even is not sharp.

\begin{figure}[t]
\begin{minipage}{7cm}
\begin{center}
\begin{pspicture}(-1,-1)(7,3)
\psframe(.8,-.7)(5.2,2.7)
\rput[r](.7,2.5){$1$}\rput[r](.7,-.5){$n_1$}
\rput[b](1,2.85){$1$}\rput[b](2.85,2.85){${n_2\over2}$}\rput[b](5,2.85){$n_2$}
{\psset{fillstyle=solid}
\psframe[fillcolor=lightgray](.8,2.7)(2.7,2.4)
\psframe[fillcolor=darkgray](3,2.4)(5.2,-.7)
}
\psline(2.7,2.7)(2.7,-.7)\psline(3,2.7)(3,-.7)
\psline(.8,2.4)(5.2,2.4)
\psdot(2.85,2.5)\psdot(2.85,-.5)
\end{pspicture}
\end{center}
\caption{Minimum spread for the column $n_2/2$ occurs when $i_1=1$.
The elements in the the light gray area are less than the minimum in
the column, and the elements in the dark gray area are greater than the
maximum.}
\label{even_even_first_half}
\end{minipage}
\hfill
\begin{minipage}{7cm}
\begin{center}
\begin{pspicture}(-1,-1)(7,3)
\psframe(.8,-.7)(5.2,2.7)
\rput[r](.7,2.5){$1$}\rput[r](.7,-.5){$n_1$}
\rput[b](1,2.85){$1$}\rput[b](3.15,2.85){${n_2\over2}+1$}\rput[b](5,2.85){$n_2$}
{\psset{fillstyle=solid}
\psframe[fillcolor=lightgray](.8,2.7)(3,-.4)
\psframe[fillcolor=darkgray](3.3,-.4)(5.2,-.7)
}
\psline(3.3,2.7)(3.3,-.7)\psline(3,2.7)(3,-.7)
\psline(.8,-.4)(5.2,-.4)
\psdot(3.15,2.5)\psdot(3.1,-.5)
\end{pspicture}
\end{center}
\caption{Minimum spread for the column $n_2/2+1$ occurs when $i_1=n_1$.
The elements in the the light gray area are less than the minimum in
the column, and the elements in the dark gray area are greater than the
maximum.}
\label{even_even_second_half}
\end{minipage}
\end{figure}
\begin{figure}
\begin{minipage}{7cm}
\begin{center}
\begin{pspicture}(-1,-1)(7,3)
\psframe(.8,-.7)(5.2,2.7)
\rput[r](.7,2.5){$1$}\rput[r](.7,-.5){$n_1$}
\rput[b](1,2.85){$1$}\rput[b](5,2.85){$n_2$}
\psline(.8,2.4)(5.2,2.4)\psline(.8,-.4)(5.2,-.4)
\psline(3,-.7)(3,2.7)%\psline(2.7,-.7)(2.7,2.7)\psline(3.3,-.7)(3.3,2.7)
\rput(1.9,2.55){\bf 1}\rput(1.9,1){\bf 2}\rput(1.9,-.55){\bf 4}
\rput(4.1,2.55){\bf 3}\rput(4.1,1){\bf 5}\rput(4.1,-.55){\bf 6}
\end{pspicture}
\end{center}
\caption{Arrangements that maintain the minimum spreads in the the two
central columns. The areas are filled monotonically in the order shown
in the figure.}
\label{even_even_extreme}
\end{minipage}
\hfill
\begin{minipage}{7cm}
\begin{center}
\begin{pspicture}(-1,-1)(7,3)
\psframe(.8,-.7)(5.2,2.7)
\rput[r](.7,2.5){$1$}\rput[r](.7,-.5){$n_1$}
\rput[r](.7,1.3){$i_1$}\rput[r](.7,.75){$n_1-i_1$}
\rput[b](1,2.85){$1$}\rput[b](5,2.85){$n_2$}
\psline(.8,1.5)(5.2,1.5)\psline(.8,.5)(5.2,.5)
\psline(3,-.7)(3,2.7)%\psline(2.7,-.7)(2.7,2.7)\psline(3.3,-.7)(3.3,2.7)
\rput(1.9,2.15){\bf 1}\rput(1.9,1){\bf 2}\rput(1.9,-.15){\bf 4}
\rput(4.1,2.15){\bf 3}\rput(4.1,1){\bf 5}\rput(4.1,-.15){\bf 6}
\end{pspicture}
\end{center}
\caption{Arrangements that maintain the spreads achieved for first
coordinate $i_1$ in column $n_2/2$ and first coordinate $n_1-i_1$ in
column $n_2/2+1$. Since the spread in a line is a symmetric unimodal
function with the maximum for the middle coordinate, those spreads are
equal.}
\label{even_even_i}
\end{minipage}
\end{figure}
The minimum spread for the column $n_2/2$ is achieved when $i_1=1$,
while the minimum spread for the column $n_2/2+1$ is achieved when
$i_1=n_1$. All the arrangements consistent with both spreads have
the form shown in Figure \ref{even_even_extreme}. However, it is not
difficult to see that for any arrangement of this type the spread in
any row $i_1$ passing through the areas $2$ and $5$ the spread is at
least
\[(n_1-i_1){n_2\over2}+(i_1+1){n_2\over2}-1 = {(n_1+1)n_2\over2}-1.\]
The spread in rows passing through areas $1$ and $3$ or $4$ and $6$ is
at most the spread in any column, which is at least
\[{n_1n_2\over2}-1 < {(n_1+1)n_2\over2}-1.\]

Similarly, for any first coordinate $i_1$, Figure \ref{even_even_i}
shows all the arrangements consistent with the spread achieved in the
column $n_2/2$ with the first coordinate being $i_1$ and the column
$n_2/2+1$ with the first coordinate being $n_1-i_1$. Again, for any row
passing through the areas $2$ and $5$ is at least
${(n_1+1)n_2\over2}-1$.  The spread in columns is at least
\begin{eqnarray*}
(n_1-i_1){n_2\over2}+i_1({n_2\over2}+1)-1 &=& {n_1n_2\over2}+i_1-1\\
&\le& {(n_1+1)n_2\over2}-1.
\end{eqnarray*}
The best spread is achieved when there are no areas $2$ and $5$, that
is, $i_1=n_1/2$. The row spread now is at most the column spread for
any row and the column spread is at least
\begin{eqnarray*}
{n_1n_2\over2}+{n_1\over2}-1 &=& {n_1(n_2+1)\over2}-1\\
&\le& {(n_1+1)n_2\over2}-1.
\end{eqnarray*}

Thus, for any monotonic arrangement with both $n_1$ and $n_2$ even the
spread must be at least
\[{n_1(n_2+1)\over2}-1.\]
\end{proof}

We now present an arrangement that achieves this lower bound and is
thus optimal. The algorithm is slightly different for odd and even
$n_1$ therefore we will state them separately.

\begin{theorem}
\label{two_dim_upper_even}
The following algorithm produces an arrangement of spread
\[{n_1(n_2+1)\over2}-1\] if $n_1 \le n_2$ and $n_1$ is even and is
thus optimal:

Fill consecutively, left to right, the upper half-columns of the
matrix, then fill the lower half-columns of the matrix in the same
manner:
\begin{enumerate}
\item Fill consecutively, column by column, the upper half of each column $i_2$.\\
That is, fill the cells \[(1,i_2), \; (2,i_2), \; \dots, \; ({n_1\over2}, i_2)\] with
numbers \[(i_2-1) {n_1\over2}+1, (i_2-1) {n_1\over2}+2,\dots, i_2 {n_1\over2}.\]
\item Fill consecutively, column by column, the lower half of each column $i_2$.\\
That is, fill the cells 
\[({n_1\over2}+1,i_2), \; ({n_1\over2}+2,i_2),\; \dots, \; (n_1,i_2)\]
with numbers
\[{n_1n_2\over2} +(i_2-1) {n_1\over2}+1, \; {n_1n_2\over2} +(i_2-1) {n_1\over2}+2,\; \dots, \; {n_1n_2\over2} +i_2 {n_1\over2}.\]
\end{enumerate}
The arrangement is shown schematically in Figure \ref{arrangement_even}.
\end{theorem}

The proof is simply an algebraic verification of the spread in all the
rows and columns.
\begin{proof}
Since all the numbers in the upper half are less than all the values
in the lower half of the arrangement, the spread in any row is at most
the spread in any column.

The spread in any column $i_2$ is the difference between the elements
in the last row and the first row of the column:
\[\left({n_1n_2\over 2} +i_2{n_1\over2}\right) - (i_2-1){n_1\over2}-1 = 
{n_1n_2\over 2} + {n_1\over2} - 1 = 
{n_1(n_2+1)\over 2} - 1.\]

Thus the overall spread of the arrangement is ${n_1(n_2+1)/ 2} - 1$,
which is the lower bound for the case of $n_1$ even, and the
arrangement is optimal.
\end{proof}

\begin{figure}
\begin{minipage}[b]{7cm}
\begin{center}
\begin{pspicture}(-1,-1)(7,4)
\rput[r](.7,2.5){$1$}\rput[r](.7,-.5){$n_1$}
\rput[b](1,2.85){$1$}\rput[b](5,2.85){$n_2$}
\psline(.8,1)(5.2,1)
\psline[linewidth=.5pt]{->}(1,2.5)(1,1.2)
\psline[linewidth=.5pt]{<-}(1,-.5)(1,.8)
\psset{framesep=0pt}
\rput*(1,2){\small\bf 1}\rput*(1.4,2){\small\bf 2}\rput*(5,2){\small\bf 11}
\rput*(1,0){\small\bf 12}\rput*(1.4,0){\small\bf 13}\rput*(5,0){\small\bf 22}
\psframe(.8,-.7)(5.2,2.7)
\multido{\n=1.2+.4, \r=1.4+.4}{10}{
\psline(\n,2.7)(\n,-.7)
\psline[linewidth=.5pt]{->}(\r,2.5)(\r,1.2)
\psline[linewidth=.5pt]{<-}(\r,-.5)(\r,.8)
}
\end{pspicture}
\end{center}
\caption{The arrangement in case of $n_1$ even.}
\label{arrangement_even}
\end{minipage}
\hfill
\begin{minipage}[b]{7cm}
\begin{center}
\begin{pspicture}(-1,-1)(7,4)
\psframe(.8,-.7)(5.2,2.7)
\rput[r](.7,2.5){$1$}\rput[r](.7,-.5){$n_1$}
\rput[b](1,2.85){$1$}\rput[b](5,2.85){$n_2$}
\psline(.8,1.2)(5.2,1.2)\psline(.8,.8)(5.2,.8)\psline(3.2,1.2)(3.2,.8)
{\psset{linewidth=.5pt}
\psline{->}(1,2.5)(1,1.4)\psline{<-}(1,-.5)(1,.6)
\psline{->}(1,1)(3,1)\psline{->}(3.4,1)(5,1)
}
\psset{framesep=0pt}
\rput*(1,2){\small\bf 1}\rput*(1.4,2){\small\bf 2}\rput*(3,2){\small\bf 6}
\rput*(1.5,1){\small\bf 7}\rput*(3.4,2){\small\bf 8}\rput*(5,2){\small\bf 12}
\rput*(1,0){\small\bf 13}\rput*(1.4,0){\small\bf 14}\rput*(3,0){\small\bf 18}
\rput*(4,1){\small\bf 19}\rput*(3.4,0){\small\bf 20}\rput*(5,0){\small\bf 24}
\psframe(.8,-.7)(5.2,2.7)
\multido{\n=1.2+.4, \r=1.4+.4}{10}{
\psline(\n,2.7)(\n,1.2)\psline(\n,.8)(\n,-.7)
\psline[linewidth=.5pt]{->}(\r,2.5)(\r,1.4)
\psline[linewidth=.5pt]{<-}(\r,-.5)(\r,.6)
}
\end{pspicture}
\end{center}
\caption{The arrangement in case of $n_1$ odd.}
\label{arrangement_odd}
\end{minipage}
\end{figure}

\begin{theorem}
The following algorithm produces an arrangement of spread
\[{(n_1+1)n_2\over2}-1\] if $n_1 \le n_2$ and $n_1$ is odd and is
thus optimal:
\begin{enumerate}
\item Fill consecutively, column by column, the upper $\lfloor n_1/2\rfloor$
cells of columns $1$ through $\lceil n_2/2\rceil$.\\
That is, fill the cells
\[(1,i_2), \; (2,i_2), \; \dots, \; (\left\lfloor{n_1\over2}\right\rfloor, i_2)\]
with numbers
\[(i_2-1) \left\lfloor{n_1\over2}\right\rfloor+1, \; (i_2-1) \left\lfloor{n_1\over2}\right\rfloor+2,\; \dots, \; i_2\left\lfloor{n_1\over2}\right\rfloor.\]
\item Fill consecutively the left $\lceil n_2/2\rceil$ cells of the 
row $(\lfloor n_1/2\rfloor+1)=\lceil n_1/2 \rceil$.\\
That is, fill the cells
\[(\left\lceil{n_1\over2}\right\rceil,1),\; (\left\lceil{n_1\over2}\right\rceil,2),\; \dots, \; (\left\lceil{n_1\over2}\right\rceil,\left\lceil{n_2\over2}\right\rceil)\]
with numbers
\[\left\lceil{n_2\over2}\right\rceil\left\lfloor{n_1\over2}\right\rfloor+1, \; \left\lceil{n_2\over2}\right\rceil\left\lfloor{n_1\over2}\right\rfloor +2,\;
\dots, \; \left\lceil{n_2\over2}\right\rceil\left(\left\lfloor{n_1\over2}\right\rfloor+1\right).\]
\item Fill consecutively, column by column, the upper $\lfloor n_1/2\rfloor$
cells of columns $\lceil n_2/2\rceil+1$ through $n_2$.\\
That is, fill the cells
\[(1,i_2), \; (2,i_2), \; \dots, \; (\left\lfloor{n_1\over2}\right\rfloor, i_2)\]
with numbers 
\[(i_2-1)\left\lfloor{n_1\over2}\right\rfloor+\left\lceil{n_2\over2}\right\rceil+1, \; (i_2-1)\left\lfloor{n_1\over2}\right\rfloor+\left\lceil{n_2\over2}\right\rceil+2, \; \dots, \; i_2\left\lfloor{n_1\over2}\right\rfloor+\left\lceil{n_2\over2}\right\rceil.\]
\item Fill consecutively, column by column, the lower $\lfloor n_1/2\rfloor$
cells of columns $1$ through $\lceil n_2/2\rceil$.\\
That is, fill the cells 
\[(\left\lceil{n_1\over2}\right\rceil+1,i_2), \; (\left\lceil{n_1\over2}\right\rceil+2,i_2),\; \dots, \; (n_1,i_2)\]
with numbers
\[(i_2-1) \left\lfloor{n_1\over2}\right\rfloor+\left\lfloor{n_1\over2}\right\rfloor n_2 +\left\lceil{n_2\over2}\right\rceil+1, \; (i_2-1) \left\lfloor{n_1\over2}\right\rfloor+\left\lfloor{n_1\over2}\right\rfloor n_2 +\left\lceil{n_2\over2}\right\rceil+2,\; \dots, \; i_2\left\lfloor{n_1\over2}\right\rfloor+\left\lfloor{n_1\over2}\right\rfloor n_2 +\left\lceil{n_2\over2}\right\rceil.\]
\item Fill consecutively the right $\lfloor n_2/2\rfloor$ cells of the
row $\lceil n_1/2\rceil$.\\
That is, fill the cells
\[(\left\lceil{n_1\over2}\right\rceil,\left\lceil{n_2\over2}\right\rceil+1),\; 
(\left\lceil{n_1\over2}\right\rceil,\left\lceil{n_2\over2}\right\rceil+2),
\; \dots, \; (\left\lceil{n_1\over2}\right\rceil, n_2)\]
with numbers
\[\left(n_2+\left\lceil{n_2\over2}\right\rceil\right)\left\lceil{n_1\over2}\right\rceil+\left\lceil{n_2\over2}\right\rceil+1, \; \left(n_2+\left\lceil{n_2\over2}\right\rceil\right)\left\lceil{n_1\over2}\right\rceil+\left\lceil{n_2\over2}\right\rceil+2, \; \dots, \; \left(n_2+\left\lceil{n_2\over2}\right\rceil\right)\left\lceil{n_1\over2}\right\rceil+n_2.\]
\item Fill consecutively, column by column, the lower $\lfloor n_1/2\rfloor$
cells of columns $\lceil n_2/2\rceil+1$ through $n_2$.\\
That is, fill the cells
\[(\left\lceil{n_1\over2}\right\rceil+1,i_2), \; (\left\lceil{n_1\over2}\right\rceil+2,i_2),\; \dots, \; (n_1,i_2)\]
with numbers
\[(n_2+i_2-1)\left\lfloor{n_1\over2}\right\rfloor+n_2+1, \; (n_2+i_2-1)\left\lfloor{n_1\over2}\right\rfloor+n_2+2,\; \dots, \; (n_2+i_2)\left\lfloor{n_1\over2}\right\rfloor+n_2.\]
\end{enumerate}
The arrangement is shown schematically in Figure \ref{arrangement_odd}.
\end{theorem}

The proof is algebraic and is similar to the case of $n_1$ even.
\begin{proof}
The difference between the largest and the smallest number in columns
$1$ through $\lceil n_2/2\rceil$ is the difference between the
elements in the last and first rows of that column $i_2$:
\begin{eqnarray*}
\left(i_2\left\lfloor{n_1\over2}\right\rfloor+\left\lfloor{n_1\over2}\right\rfloor n_2 +\left\lceil{n_2\over2}\right\rceil\right) -
(i_2 -1)\left\lfloor{n_1\over2}\right\rfloor -1 &=& 
\left\lfloor{n_1\over2}\right\rfloor n_2+\left\lfloor{n_1\over2}\right\rfloor
+\left\lceil{n_2\over2}\right\rceil - 1\\
&<&
\left\lfloor{n_1\over2}\right\rfloor n_2+\left\lfloor{n_2\over2}\right\rfloor
+\left\lceil{n_2\over2}\right\rceil - 1\\
&=& {(n_1+1)n_2\over2}-1,
\end{eqnarray*}
since $n_1\le n_2$.

The difference between the largest and the smallest number in columns
$\lceil n_2/2\rceil+1$ through $n_2$ is, again, the difference between
the elements in the last and first rows of that column $i_2$:
\[
\left((n_2+i_2)\left\lfloor{n_1\over2}\right\rfloor+n_2\right) - 
\left((i_2-1)\left\lfloor{n_1\over2}\right\rfloor+\left\lceil{n_2\over2}\right\rceil+1\right) = 
\left\lfloor{n_1\over2}\right\rfloor n_2+\left\lfloor{n_1\over2}\right\rfloor
+\left\lceil{n_2\over2}\right\rceil - 1,
\]
which is the same as in the other columns, and thus less than $(n_1+1)n_2/2-1.$

It is easy to see that the largest spread in any row is achieved in
row $\lfloor n_1/2\rfloor$. The difference between the largest and the
smallest number in that row is
\[
\left(\left(n_2+\left\lceil{n_2\over2}\right\rceil\right)\left\lfloor{n_1\over2}\right\rfloor+n_2\right) - 
\left\lceil{n_2\over2}\right\rceil\left\lfloor{n_1\over2}\right\rfloor-1 =
n_2\left(\left\lfloor{n_1\over2}\right\rfloor+1\right)-1 =
{(n_1+1)n_2\over2}-1.
\]

Thus the overall spread of the arrangement is ${(n_1+1)n_2/ 2} - 1$,
which is the lower bound for the case of $n_1$ odd, and so the
arrangement is optimal.
\end{proof}

We have shown that the proposed algorithm produces an arrangement with
the spread that matches the lower bound of
\[{(n_1+1)n_2\over 2} - 1 \mbox{ if }n_1\mbox{ is odd, }\]
\[{n_1(n_2+1)\over 2} - 1 \mbox{ if }n_1\mbox{ is even},\]
and thus is optimal.

%-----------------------------k D----------------------------

\subsection{Generalization to Arbitrary Dimensions\\}
Given a $n_1\times n_2\times\cdots\times n_d$ matrix, where $n_1 \le n_2
\le\cdots\le n_d$, the goal is to arrange the numbers $\{1,\dots,
\prod{n_t}\}$ in a way that minimizes the maximum difference between
the largest and the smallest number in any line of the matrix.

Using techniques very similar to the two-dimensional case, it is
possible to show a lower bound of roughly
\[B(K_2^d)\times\prod{n_t\over2},\] where $B(K_2^d)$ is the bandwidth
of the product of $d$ $2$-cliques, and give an arrangement that nearly
achieves it.

\begin{theorem}
The spread in any arrangement of an $n_1\times\cdots\times n_d$
matrix, where $n_1 \le n_2\le\cdots\le n_d$, is at least
\[	B(K_2^d) \times \prod_{t=1}^d{\left\lfloor{n_t\over 2}\right\rfloor}.
\]
\end{theorem}

\begin{proof}
By Lemma \ref{monotonic_lower} the spread in the optimal arrangement
is at least
\[\max_{\mbox{\footnotesize all lines}}{\{
    \min_{\mbox{\footnotesize cells in a line}}{\{
      \min_{\mbox{\footnotesize separating set }\atop
	    \mbox{\footnotesize of }B(K_2^d)\mbox{ \footnotesize orthants}}{\{
        \mbox{volume of the separating set of orthants}
      \}}
    \}}
  \}}.
\]
Just like in two dimensions, the smallest volume of the separating
orthants for any line $(i_1,...,*,...,i_d)$ occurs either for the cell
$i_j=1$ or $i_j=n_j$, depending on which of the coordinates $i_t,
t\neq j$ are at most $\lceil n_t/2\rceil$ and which ones are
greater. Thus the maximum over all lines of the minimum volume of the
separating orthants is achieved for one of the extreme cells of the
central lines
$(\lceil n_1/2\rceil, \lceil n_2/2\rceil,...,*,...,\lceil n_d/2\rceil)$.
This in itself immediately gives a lower bound of
\[	B(K_2^d) \times \prod_{t=1}^d{\left\lfloor{n_t\over 2}\right\rfloor}.
\]
\end{proof}

The optimal arrangement in $d$ dimensions is constructed similarly to
the $2$-dimensional one.

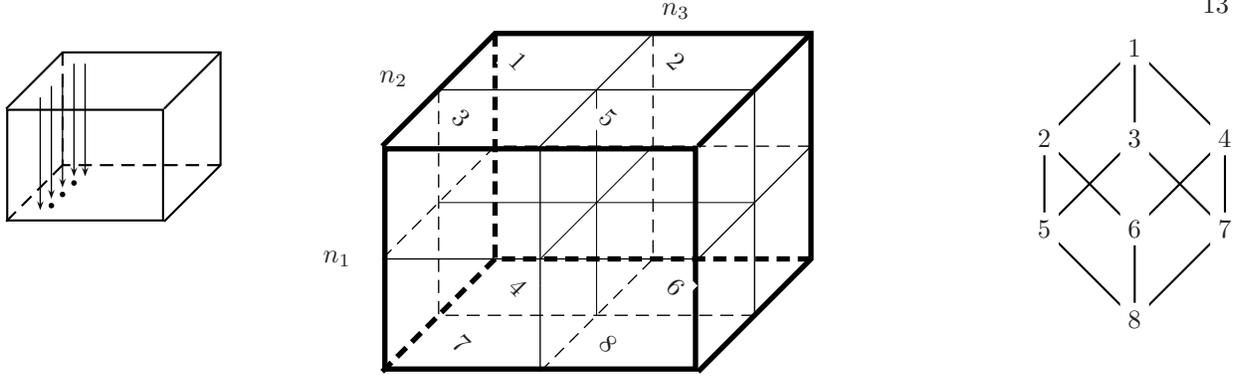
\begin{figure}
\begin{center}
\begin{pspicture}(-5,-2)(5,3)
\rput(-8,0){\psset{unit=.3}
\psframe(0,0)(7,5)
\psline(0,5)(2.5,7.5)(9.5,7.5)(9.5,2.5)(7,0)\psline(7,5)(9.5,7.5)
\psset{linestyle=dashed}
\psline(2.5,7.5)(2.5,2.5)(9.5,2.5)
\psline(0,0)(2.5,2.5)
\psset{linewidth=.5pt,linestyle=solid}
\multirput(3,7)(-.5,-.5){4}{\psline{->}(0,0)(0,-5)}
\psline{->}(3.5,7)(3.5,2)
\multirput(3,1.7)(-.5,-.5){3}{\psdot[linewidth=.05pt](0,0)}
}
\rput(-3,-2){\psset{unit=.6}
\psset{linewidth=2pt}
\psframe(0,0)(7,5)
\psline(0,5)(2.5,7.5)(9.5,7.5)(9.5,2.5)(7,0)\psline(7,5)(9.5,7.5)
\psset{linestyle=dashed}
\psline(2.5,7.5)(2.5,2.5)(9.5,2.5)
\psline(0,0)(2.5,2.5)
\psset{linewidth=.5pt, linestyle=solid}
\psline(3.5,0)(3.5,5)(6,7.5)
\psline(8.25,1.25)(8.25,6.25)(1.25,6.25)
\psline(0,2.5)(7,2.5)(9.5,5)
\psset{linestyle=dashed}
\psline(6,7.5)(6,2.5)(3.5,0)
\psline(1.25,6.25)(1.25,1.25)(8.25,1.25)
\psline(9.5,5)(2.5,5)(0,2.5)
\psset{linewidth=.1pt, linestyle=solid}
\psline(4.75,1.25)(4.75,6.25)
\psline(1.25,3.75)(8.25,3.75)
\psline(3.5,2.5)(6,5)
\rput*{-45}(3,6.875){$1$}\rput*{-45}(6.5,6.875){$2$}
\rput*{-45}(1.75,5.625){$3$}\rput*{-45}(3,1.875){$4$}
\rput*{-45}(5,5.625){$5$}\rput*{-45}(6.5,1.875){$6$}
\rput*{-45}(1.75,.625){$7$}\rput*{-45}(5,.625){$8$}
\rput(-1,2.5){$n_1$}\rput(.25,6.5){$n_2$}\rput(6.5,8){$n_3$}
}
\rput(7,.5){\psset{unit=.6}
\psset{linewidth=.8pt}
\psline(2,-1)(2,1)(0,3)(-2,1)(-2,-1)(0,-3)(2,-1)(0,1)(-2,-1)
\psline(-2,1)(0,-1)(2,1)\psline(0,-1)(0,-3)\psline(0,1)(0,3)
\rput*(0,3){$1$}
\rput*(-2,1){$2$}\rput*(0,1){$3$}\rput*(2,1){$4$}
\rput*(-2,-1){$5$}\rput*(0,-1){$6$}\rput*(2,-1){$7$}
\rput*(0,-3){$8$}
}
\end{pspicture}
\end{center}
\caption{Schematic representation of $3$-dimensional arrangement in case of
$n_1$ even.}
\label{alg_even3}
\end{figure}

\begin{theorem}
The following algorithm produces an arrangement $A$ of spread at most
\[B(K_2^d) \times\prod{\left\lceil{n_t\over2}\right\rceil}+{n_1\over2}-1\]
in case $n_1=\min_t{\{n_t\}}$ is even and is thus nearly optimal:
\begin{enumerate}
\item Divide the matrix into $2^d$ orthants by dividing each coordinate
$n_t$ into two halves of size $\lfloor{n_t/2}\rfloor$ and $\lceil{n_t/2}\rceil$
\item Fill the first orthant (containing the coordinate $(1,...,1)$) in
the following way:
	\begin{enumerate}
	\item $A(1) = (1,...,1)$
	\item The $1$st coordinate of $A(m)$ is the $1$st coordinate of
	$A(m-1)$ plus $1$ modulo $\lfloor{n_1/2}\rfloor$.\\  If the $t$th
	coordinate becomes $1$ then the $(t+1)$st coordinate increases
	by $1$ modulo $\lfloor{n_{t+1}/2}\rfloor$.
	\end{enumerate}
\item Fill the orthants one after another in a way similar to the first
orthant. The orthants are filled in the order corresponding to the
optimal numbering of $K_2^d$ \cite{harper2}: at each step number a
neighbor of the smallest already numbered vertex, taking care that the
maximum bandwidth difference occurs between the vertices in $K_2^d$
adjacent along the $n_1$ coordinate. To ensure that, after numbering
the vertex corresponding to the first orthant with number $1$, number
the orthant adjacent to it along the $(d-t+1)$th coordinate with
number $t$.
\end{enumerate}
\end{theorem}
The algorithm is shown schematically for $3$ dimensions in Figure
\ref{alg_even3}.
\begin{proof}
Any line in the arrangement is contained within two orthants,
therefore the spread in any line is the difference between the labels
of the corresponding orthants times the volume of the larger-volume
orthant plus the difference between the smallest and the largest
number of the line within the smaller-volume orthant. Notice that for
any two orthants with the label difference less than the bandwidth of
the $d$-dimensional hypercube the spread in the line passing through
them is at most the bandwidth times the volume of the larger-volume
orthant. Therefore the maximum spread occurs in a line passing through
two orthants with the label difference equal to the bandwidth of the
$d$-dimensional hypercube. By construction all such orthants align
in the direction of the first dimension, that is along the $i_1$
coordinate. Thus the spread in that line is at most
\[B(K_2^d) \times\prod{\left\lceil{n_t\over2
}\right\rceil}+{n_1\over2}-1\]
\end{proof}

\begin{theorem}
The following algorithm produces an arrangement $A$ of spread at most
\[B(K_{n_2}\times\cdots\times K_{n_d})+
B(K_2^d)\left\lfloor{n_1\over2}\right\rfloor
\prod_{t=2}{\left\lceil{n_t\over2}\right\rceil}
\]
in case $n_1=\min_t{\{n_t\}}$ is odd and is thus nearly optimal:
\begin{enumerate}
\item Divide the matrix into $2^d$ orthants by dividing each coordinate
$n_t \ne n_1$ into two halves of size $\lfloor{n_t/2}\rfloor$ and
$\lceil{n_t/2}\rceil$. For $n_1$ use the coordinates $i_1 < \lceil
n_1/2\rceil$ and $i_1 > \lceil n_1/2\rceil$ to define the rest of
the orthants. The submatrix $(\lceil n_1/2\rceil,*,*,...,*)$ is left
out.
\item Fill the first orthant (containing the coordinate $(1,...,1)$) in
the following way:
	\begin{enumerate}
	\item $A(1) = (1,...,1)$
	\item The $1$st coordinate of $A(m)$ is the $1$st coordinate of
	$A(m-1)$ plus $1$ modulo $\lfloor{n_1/2}\rfloor$.\\  If the $t$th
	coordinate becomes $1$ then the $(t+1)$st coordinate increases
	by $1$ modulo $\lfloor{n_{t+1}/2}\rfloor$.
	\end{enumerate}
\item Fill the orthants one after another in a way similar to the first
orthant. The orthants are filled in the same way and the same order as
in the case of $n_1$ even up to and including the orthant
corresponding to the first vertex on the edge in the hypercube that
gives the maximum bandwidth.
\item Recursively fill the shadow of the filled orthants in the
$d-1$-dimensional submatrix $(\lceil n_1/2\rceil,*,*,...,*)$ with
the optimal arrangement.
\item Fill the orthants up to the orthant that corresponds to
the other vertex on the first edge in the hypercube with the maximum
bandwidth.
\item Recursively fill the rest of the $d-1$-dimensional submatrix
$(\lceil n_1/2\rceil,*,*,...,*)$ with the optimal arrangement.
\item Fill the rest of the orthants.
\end{enumerate}
%The algorithm is shown schematically for $3$ dimensions in Figure
%\ref{alg_odd3}.
\end{theorem}

\begin{proof}
Similar to the case of $n_1$ even, each line passes through either two
orthants above the submatrix $(\lceil n_1/2\rceil,*,*,...,*)$, below
that submatrix, through an orthant above, an orthant below, and the
submatrix, or lies entirely within the submatrix. It is easy to see that the
maximum spread occurs in a line of the last type and is therefore, 
\[B(K_{n_2}\times\cdots\times K_{n_k})+B(K_2^d)\left\lfloor{n_1\over2}\right\rfloor\prod_{t=2}{\left\lceil{n_t\over2}\right\rceil}
\]
\end{proof}

Thus we have shown that the bandwidth of a Hamming graph
$K_{n_1}\times\cdots\times K_{n_d}$ is between the lower bound LB and
the upper bound UB, where LB and UB are as follows:
\begin{eqnarray*}
\mathbf{LB} &=&
	B(K_2^d) \times \prod_{t=1}^d{\left\lfloor{n_t\over 2}\right\rfloor}
\\\\
\mathbf{UB} &=& \left\{{\begin{array}{lr}
	\displaystyle B(K_2^d) \times\prod{\left\lceil{n_i\over2}\right\rceil} +
		{n_1\over2}-1 &\mbox{if $n_1$ is even}\\\\
	\displaystyle B(K_{n_2}\times\cdots\times K_{n_d})+
		B(K_2^d)\left\lfloor{n_1\over2}\right\rfloor
		\prod_{t=2}^d{\left\lceil{n_t\over2}\right\rceil}
		& \mbox{if $n_1$ is odd}
			 \end{array}}\right\}
\end{eqnarray*}
Notice, if all $n_t$ are even, then the difference between the LB and
UB is $n_1/2-1$, which is very small compared to the order of magnitude
of the LB of $O((n_1/2)^d)$. The difference between LB and UB is
largest when all $n_t$ are odd. Let all $n_t$ be equal $n$. Noting that
\[ B(K_{n_2}^d) = \sum_{t=0}^{d-1}{i \choose \lfloor t/2\rfloor} 
\approx {2^{d-1}\over \sqrt{d-1}},
\]
the upper bound is
\begin{eqnarray*}
\mathbf{UB}  &=& B(K_{n_2}\times\cdots\times K_{n_d})+
			B(K_2^d)\left\lfloor{n_1\over2}\right\rfloor
			\prod_{t=2}^d{\left\lceil{n_t\over2}\right\rceil}\\
&=& \sum_{r=1}^d{B(K_2^r)\lfloor{n_{d-r+1}\over2}\rfloor\prod_{t=r-2}^d{\lceil{n_{d-t}\over 2}\rceil}}\\
&=& \sum_{r=1}^d{B(K_2^r){n-1\over2}\left({(n+1)\over2}\right)^{r-1}}\\
&\approx& \sum_{r=1}^d{{2^{r-1}\over\sqrt{r-1}}{n^{r}\over 2^{r}}}\\
&\approx& \sum_{r=1}^{d}{n^r\over2\sqrt{r-1}}
\end{eqnarray*}
while the lower bound is approximately
\[\mathbf{LB} \approx {n^d\over2\sqrt{d-1}}.
\]
Thus, the difference between LB and UB in case $n$ is odd is the order
of $O(n^{d-1})$.

Therefore, overall, the upper and lower bounds nearly coincide in
infinitely many points. We believe that the upper bound is the correct
bandwidth of the Hamming graph and the lower bound needs to be
tightened.

\section{Acknowledgments}
We are deeply grateful to Lawrence Harper for giving us the idea for
the algorithm and for many constructive comments.

\end{document}